# Increase of the pion-kaon Hawking radiation from Schwarzschild black holes by Dirac monopoles

Yu. P. Goncharov

*Theoretical Group, Experimental Physics Department, State Technical University, Sankt-Petersburg 195251, Russia*

N. E. Firsova

*Institute for Problems of Mechanical Engineering, Russian Academy of Sciences, Sankt-Petersburg 199178, Russia*

**Abstract**

An algorithm for numerical computation of the barrier transparency for the potentials surrounding Schwarzschild black holes is described for massive scalar particles. It is then applied to calculate the total (including all particle species and the contributions of twisted field configurations connected with Dirac monopoles) luminosity for the pion-kaon Hawking radiation from a Schwarzschild black hole with mass $M = 10^{12}$ g. It is found that the contribution due to monopoles can be of order 11 % of the total pion-kaon luminosity.

## 1 Introductory remarks

In a number of papers [1,2] it was shown that owing to nontrivial $\mathbb{R}^2 \times \mathbb{S}^2$-topology of black hole spacetimes the topologically inequivalent configurations (TICs) of complex scalar field can exist near black holes. Physically the appearance of TICs for the given field should be obliged to the natural presence of Dirac monopoles on black hole and because of the interaction with them complex scalars splits into TICs though the total (internal) magnetic charge of black hole remains equal to zero (see for more details Refs. [1,2] and below). As a consequence, description of the Hawking radiation process should be modified to include the contributions from different TICs.

The expression for the luminosity regarding Hawking radiation for any TIC, however, contains an element of the $S$-matrix connected with some potential barrier surrounding black hole and which effectively arises for the quantum scalar particle leaving black hole. One should therefore solve some scattering problem on the whole axis with the given potential. The relevant potentials, however, depend on both the type of black hole and the particle masses and are complicated since they can be described only in an implicit form. Therefore for physical results to be obtained one needs to apply the numerical methods. In the massless case the scattering problems discussed are simpler to treat and we obtained numerical results for luminosities in the case of Schwarzschild (SW) and Reissner-Nordström (RN) black holes in Refs. [3]. The massive case, however, is more realistic but the corresponding scattering problems are more complicated. Recently they have been studied in Ref. [4] in the SW case where it was possible to find an algorithm for numerical calculation of the conforming $S$-matrix element and, accordingly, of the conforming Hawking radiation luminosity for massive scalar particles. The present paper contains a description of the algorithm and applies it to calculate the total luminosity for pions and kaons (both neutral and charged ones) for a SW black hole with mass $M = 10^{12}$ g. Of course, the given scalar mesons are not genuine elementary particles since they are composed of quarks but one can assume that the quark hadronization to some extent occurs inside black holes. Actually such an assumption is reasonable in light of a recent solution to the puzzle of the spin content of the proton, which involves assuming the proton contains virtual pions (as well as virtual quarks, antiquarks and gluons) and then the observed asymmetry can arise quite naturally (see, for instance, Ref. [5] for more details and further references). It is therefore reasonable to assume that the more massive objects, such as black holes contain virtual light mesons, such as pions and kaons. As a result, the task under consideration here makes sense.

We write down the black hole metric under discussion (using the ordinary set of local coordinates $t, r, \vartheta, \varphi$) in the form

$$ds^2 = adt^2 - a^{-1}dr^2 - r^2(d\vartheta^2 + \sin^2\vartheta d\varphi^2) \qquad (1)$$

with $a = 1 - r_g/r$, $r_g = 2M$ and $M$ is the black hole mass.

Throughout the paper we employ the system of units with $\hbar = c = G = 1$, unless explicitly stated otherwise. Also we shall denote the set of the modulo square integrable complex functions on any manifold $F$ furnished with an integration measure as $L_2(F)$. When computing we use the following values of the scalar meson masses (see Ref. [6])

$$\mu_0(\pi^0) = 134.9764 \text{ Mev}, \mu_0(\pi^\pm) = 139.56995 \text{ Mev},$$

$$\mu_0(K^0, \overline{K}^0) = 497.672 \text{ Mev}, \mu_0(K^\pm) = 493.672 \text{ Mev}. \qquad (2)$$

## 2 Description of algorithm

As was disscussed in [1,2] TICs of a complex scalar field on black holes are conditioned by the availability of a countable number of complex line bundles over the $\mathbb{R}^2 \times \mathbb{S}^2$ -topology underlying the 4D black hole physics. Each TIC corresponds to sections of a complex line bundle $E$, which can be characterized by its Chern number $n \in \mathbb{Z}$ (the set of integers). TIC with $n = 0$ can be called *untwisted*, while the rest of the TICs with $n \neq 0$ should be reffered to as *twisted*. Using the fact that all the mentioned line bundles can be trivilized over the chart of local coordinates $(t, r, \vartheta, \varphi)$ covering almost the whole manifold $\mathbb{R}^2 \times \mathbb{S}^2$ one can obtain a suitable wave equation on the given chart for TIC $\phi$ with mass $\mu_0$ and Chern number $n \in \mathbb{Z}$ in the form

$$(|g|)^{-1/2}(\partial_\mu - ieA_\mu)[g^{\mu\nu}(|g|)^{1/2}(\partial_\nu - ieA_\nu)\phi] = -\mu_0^2 \phi, \qquad (3)$$

where vector-potential for the corresponding Dirac monopole is $A = A_\mu dx^\mu = -\frac{n}{e}\cos\vartheta d\varphi$. As was shown in Refs. [1], integrating $F = dA$ over the surface $t = const$, $r = const$ with topology $\mathbb{S}^2$ gives rise to the Dirac charge quantization condition

$$\int_{\mathbb{S}^2} F = 4\pi\frac{n}{e} = 4\pi q$$

with magnetic charge $q$, so we can identify the coupling constant $e$ with electric charge. Besides, the Maxwell equations $dF = 0$, $d*F = 0$ are fulfilled [1] with the exterior differential $d = \partial_t dt + \partial_r dr + \partial_\vartheta d\vartheta + \partial_\varphi d\varphi$ in coordinates $t, r, \vartheta, \varphi$, where $*$ denotes the Hodge dual form. Also it should be emphasized that the total (internal) magnetic charge $Q_m$ of black hole which should be considered as the one summed up over all the monopoles remains equal to zero because

$$Q_m = \frac{1}{e}\sum_{n\in\mathbb{Z}} n = 0.$$

As was shown in [1,2] Eq.(3) has in $L_2(\mathbb{R}^2 \times \mathbb{S}^2)$ a complete set of solutions of the form

$$f_{\omega m l n} = \frac{1}{r}e^{i\omega t}Y_{nlm}(\vartheta, \varphi)R_{\omega l n}(r), \qquad l = |n|, |n|+1, ..., |m| \leq l, \qquad (4)$$

where the explicit form the *monopole (spherical) harmonics* $Y_{nlm}(\vartheta, \varphi)$ can be found in [1].

We now consider the functions $R_{\omega ln}(r)$. Denoting $k = r_g \omega$, $y(x) = r/r_g$ and introducing the functions $\psi(x, k, l, n) = R_{\omega(k)ln}(r(x))$, where $y(x)$ is a function reverse to $x(y) = y + \ln(y - 1)$ (i.e., $-\infty < x < \infty$, $1 \le y < \infty$), we find that $\psi(x, k, l, n)$ obeys the Schrödinger-like equation

$$\left(\frac{d^2}{dx^2} + k^2\right)\psi(x, k, l, n) = V(x, l, n, \mu)\psi(x, k, l, n), \qquad (5)$$

$$V(x, l, n, \mu) = \left[1 - \frac{1}{y(x)}\right]\left[\mu^2 + \frac{l(l+1) - n^2}{y^2(x)} + \frac{1}{y^3(x)}\right], \qquad (6)$$

where $\mu = \mu_0 r_g$, or, through usual units, if $\mu_0$ and $M$ in g then $\mu = 2\mu_0 M/\mu_{pl}^2$, while if $\mu_0$ in Mev and $M$ in g then

$$\mu = 2 \cdot 3.762426569 \cdot 10^{-18} \mu_0 M,$$

while $\mu_{pl}^2 = c\hbar/G$ is the Planck mass square in g$^2$. It is clear that as $x \to +\infty$, $V(x, l, n, \mu) \to \mu^2$ and then condition, which holds in the massless case

$$\int\limits_{-\infty}^{+\infty} |V| dx < \infty$$

is violated and we cannot pose the scattering problem for Eq. (5) as in the massless case [3]. As an example, in Fig. 1 the numerically computed potential $V$ is shown for charged kaons at $n = 2$, $M = 10^{15}$ g.

Fig. 1. Typical potential barrier for charged kaons.

As was shown in Ref. [4], the correct statement of the scattering problem for Eq. (5) with potential (6) consists in searching for two solutions $\psi^+(x, k^+, l, n)$, $\psi^-(x, k, l, n)$ of the equation (5) satisfying the following conditions

$$\psi^+(x, k^+, l, n) = \begin{cases} e^{ikx} + s_{12}(k, l, n)e^{-ikx} + o(1), & x \to -\infty, \\ s_{11}(k, l, n)w_{-i\tilde{\mu}, \frac{1}{2}}(-2ik^+x) + o(1), & x \to +\infty, \end{cases}$$

$$\psi^-(x, k, l, n) = \begin{cases} s_{22}(k, l, n)e^{-ikx} + o(1), & x \to -\infty, \\ w_{i\tilde{\mu}, \frac{1}{2}}(2ik^+x) + s_{21}(k, l, n)w_{-i\tilde{\mu}, \frac{1}{2}}(-2ik^+x) + o(1), & x \to +\infty, \end{cases}$$
$$(7)$$

where $k^+(k) = \sqrt{k^2 - \mu^2}$, $\tilde{\mu} = \frac{\mu^2}{(k^+)^2} = \frac{\mu^2}{k^2 - \mu^2}$ and the functions $w_{\pm i\tilde{\mu}, \frac{1}{2}}(\pm z)$ are related to the Whittaker functions $W_{\pm i\tilde{\mu}, \frac{1}{2}}(\pm z)$ (concerning the latter ones see e. g. Ref. [7]) by the relation

$$w_{\pm i\tilde{\mu}, \frac{1}{2}}(\pm z) = W_{\pm i\tilde{\mu}, \frac{1}{2}}(\pm z)e^{-\pi\tilde{\mu}/2},$$

so that one can easily gain asymptotics (using the corresponding ones for Whittaker functions [7])

$$w_{i\tilde{\mu},\frac{1}{2}}(-2ik^+x) = e^{ik^+x}e^{i\tilde{\mu}\ln|2k^+x|}[1 + O(|k^+x|^{-1})],\ x \to +\infty,$$
$$w_{-i\tilde{\mu},\frac{1}{2}}(2ik^+x) = e^{-ik^+x}e^{-i\tilde{\mu}\ln|2k^+x|}[1 + O(|k^+x|^{-1})],\ x \to +\infty. \quad (8)$$

We can see that there arises some $S$-matrix with elements $s_{ij}, i,j = 1,2$ and it satisfies certain unitarity relations (for more details see Ref. [4]). As can be seen, the equation (3) corresponds to the lagrangian

$$\mathcal{L} = (-g)^{1/2}(g^{\mu\nu}\overline{\mathcal{D}_\mu\phi}\mathcal{D}_\nu\phi - \mu_0^2\overline{\phi}\phi),$$

where the overbar signifies complex conjugation, $\mathcal{D}_\mu = \partial_\mu - ieA_\mu$ and, as a result, we have the conforming energy-momentum tensor for TIC with the Chern number $n$

$$T_{\mu\nu} = \mathrm{Re}[(\overline{\mathcal{D}_\mu\phi})(\mathcal{D}_\nu\phi) - \frac{1}{2}g_{\mu\nu}g^{\alpha\beta}(\overline{\mathcal{D}_\alpha\phi})(\mathcal{D}_\beta\phi) - \mu_0^2 g_{\mu\nu}\overline{\phi}\phi].$$

Having obtained these relations, one can discuss the Hawking radiation process for any TIC of complex scalar field. Actually, one should use the above energy-momentum tensor to find, according to the standard prescription (see Refs. [1,2] and cited therein) and employing the asymptotics (7)–(8), the luminosity $L(n)$ of Hawking radiation for TIC of a massive complex scalar field with Chern number $n$ and mass $\mu_0$ is (in ordinary units)

$$L(n) = \lim_{r\to\infty} 4\pi r^2 <T_{tr}> = A\sum_{l=|n|}^{\infty}(2l+1)c_l(n),\ c_l(n) = \int_\mu^\infty \frac{\Gamma(k,l,n)k^+ dk}{e^{8\pi k^+} - 1} \quad (9)$$

with the vacuum expectation value $<T_{tr}>$, where $A = \frac{1}{2\pi\hbar}\left(\frac{\hbar c^3}{GM}\right)^2 \approx 0.273673 \cdot 10^{50}\,\mathrm{erg\cdot s^{-1}} \cdot M^{-2}$ ($M$ in g), while the barrier transparency $\Gamma(k,l,n) = |s_{11}(k,l,n)|^2$.

It is clear that for the all configurations luminosity $L$ of a black hole with respect to the Hawking radiation concerning the complex scalar field with a mass $\mu_0$ to be obtained one should sum up over all $n$

$$L = \sum_{n\in\mathbb{Z}} L(n) = L(0) + 2\sum_{n=1}^{\infty} L(n), \quad (10)$$

since $L(-n) = L(n)$. Obviously, for neutral scalar field there exists only $L(0)$ since it does not interact with monopoles. Also it is evident that for numerical computation of all the above luminosities one needs to have some algorithm for calculating $s_{11}(k,l,n)$. This can be extracted from the results of Ref. [4]. Namely
$$s_{11}(k,l,n) = 2ik/[f^-(x,k,l,n), f^+(x,k^+,l,n)], \quad (11)$$
where [,] signifies the Wronskian of functions $f^-, f^+$, the so-called Jost type solutions of Eq. 5. In their turn, these functions and their derivatives obey the following integral equations

$$f^-(x,k,l,n) = e^{-ikx} + \frac{1}{k}\int_{-\infty}^{x}\sin[k(x-t)]V(t,l,n,\mu)f^-(t,k,l,n)dt, \quad (12)$$

$$(f^-)'_x(x,k,l,n) = -ike^{-ikx} + \int_{-\infty}^{x}\cos[k(x-t)]V(t,l,n,\mu)f^-(t,k,l,n)dt, \quad (13)$$

$$f^+(x,k^+,l,n) = w_{i\tilde{\mu},\frac{1}{2}}(-2ik^+x) +$$
$$\frac{1}{k^+}\int_{x}^{+\infty}\text{Im}[w_{i\tilde{\mu},\frac{1}{2}}(-2ik^+x)w_{-i\tilde{\mu},\frac{1}{2}}(2ik^+t)]V^+(t,l,n,\mu)f^+(t,k^+,l,n)dt, \quad (14)$$

$$(f^+)'_x(x,k^+,l,n) = \frac{d}{dx}w_{i\tilde{\mu},\frac{1}{2}}(-2ik^+x) +$$
$$\frac{1}{k^+}\int_{x}^{+\infty}\text{Im}[\frac{d}{dx}w_{i\tilde{\mu},\frac{1}{2}}(-2ik^+x)w_{-i\tilde{\mu},\frac{1}{2}}(2ik^+t)]V^+(t,l,n,\mu)f^+(t,k^+,l,n)dt, \quad (15)$$

where the potential
$$V^+(x,l,n,\mu) = \mu^2\left[\frac{1}{x} - \frac{1}{y(x)}\right] + \left[1 - \frac{1}{y(x)}\right]\left[\frac{l(l+1) - n^2}{y^2(x)} + \frac{1}{y^3(x)}\right]$$

tends to 0 when $x \to +\infty$. Eqs. (12)–(13) are suitable for direct calculation while this is not the case for Eqs. (14)–(15) — it is difficult to evaluate the functions $w(z)$ in (14)–(15) since there is no effective method to compute the Whittaker functions $W(z)$ at $z$ required. In the present paper we, therefore, stick to the following strategy. We employ the asymptotic expressions of $w$-functions at $x \to +\infty$ from Eq. (8) and replace Eqs. (14)–(15) by

$$f^+(x,k^+,l,n) = e^{i(k^+x+\tilde{\mu}\ln(2k^+x))} +$$
$$\frac{1}{k^+}\int_{x}^{+\infty}\text{Im}[e^{i(k^+(x-t)+\tilde{\mu}\ln(x/t))}]V^+(t,l,n,\mu)f^+(t,k^+,l,n)dt, \quad (16)$$

$$(f^+)'_x(x,k^+,l,n) = ie^{i(k^+x+\tilde{\mu}\ln(2k^+x))}(k^+ + \tilde{\mu}/x) +$$
$$\int_{x}^{+\infty}\text{Im}[i(1+\frac{\tilde{\mu}}{k^+x})e^{i(k^+(x-t)+\tilde{\mu}\ln(x/t))}]V^+(t,l,n,\mu)f^+(t,k^+,l,n)dt. \quad (17)$$

The latter equations are appropriate for numerical evaluation but under this approximation, as numerical experiment shows, reliable calculations are only possible for $\mu < 0.01$. This upper limit corresponds to $\mu_0(\pi^\pm)$ and $M \sim 10^{13}$ g, so that for kaons the calculations are unreliable for $M = 10^{13}$ g. Therefore in the present paper we restrict ourselves to considering pions and kaons for $M = 10^{12}$ g. Including the heavier scalar mesons is only possible for smaller $M$.

Finally, one should touch upon the convergence of the series (10) over $n$. For this aim we represent the coefficients $c_l(n)$ of (9) in the form (omitting the integrand)

$$c_l(n) = c_{l1}(n) + c_{l2}(n) = \int_\mu^{(\ln l)/2\pi} + \int_{(\ln l)/2\pi}^\infty , \qquad (18)$$

so that $L(n)$ of (9) is equal to $L_1(n) + L_2(n)$ respectively. Also it should be noted, as follows from the results of Ref. [4], we have for barrier transparency $\Gamma(k, l, n)$ of (9) that $\Gamma(\mu, l, n) = 0$ and at $k \to +\infty$

$$\Gamma(k, l, n) = |s_{11}(k, l, n)|^2 = 1 + O(k^{-1}) . \qquad (19)$$

Further the asymptotic behaviour for large $l$ and $k << l$

$$|s_{11}(k, l, n)| = \frac{e^{-l}}{l}(k - \mu)[1 + o(1)] \qquad (20)$$

is crucial. To obtain it we should work using the equations (11)–(15) by conventional methods of mathematical analysis so that the detailed derivation of (20) lies somewhat out of the scope of the present paper and will be considered elesewhere. Using (20) we have the following chain of inequalities

$$c_{l1}(n) = \int_\mu^{(\ln l)/2\pi} \frac{|s_{11}|^2 k^+ dk}{e^{8\pi k^+} - 1} \leq B_1 \frac{e^{-2l}}{l^2} \int_\mu^{(\ln l)/2\pi} k^3 dk \leq B_2 e^{-2l} \frac{\ln^4 l}{l^2} \leq B_3 e^{-2l}$$

with some constants $B_i = B_i(\mu)$ depending on $\mu$ so that $(2l+1)c_{l1}(n) \leq Ble^{-2l}$ with some constant $B = B(\mu)$ depending on $\mu$. Since

$$\int_{l=|n|}^\infty le^{-2l}dl = (|n| + 1/2)e^{-2|n|}/2$$

then according to the Cauchy theorem

$$L_1(n) = A \sum_{l=|n|}^{\infty} (2l+1)c_{l1}(n) \leq AB(|n|+1/2)e^{-2|n|} . \tag{21}$$

At the same time due to (19)

$$c_{l2}(n) \leq D \int_{(\ln l)/2\pi}^{\infty} k^+ e^{-8\pi k^+} dk \sim D\frac{\ln l}{l^4} \tag{22}$$

with some constant $D = D(\mu)$ depending on $\mu$. This entails

$$L_2(n) \sim AD \int_{|n|}^{\infty} \frac{(2l+1)\ln l}{l^4} dl \sim AD\frac{\ln |n|}{n^2} . \tag{23}$$

Under this situation, according to (21) and (23), we obtain that $L(n) \sim L_2(n)$ and the series of (10) is convergent, i. e., $L < +\infty$. Consequently, we can say that all the further computed luminosities are well defined and exist.

## 3  Numerical results

In view of (19), we have [denoting the Riemann zeta function as $\zeta(s)$]

$$\int_{\mu}^{\infty} \frac{\Gamma(k,l,n)k^+ dk}{e^{8\pi k^+}-1} \sim \int_{0}^{\infty} \frac{kdk}{e^{8\pi k}-1} = \frac{1}{(8\pi)^2}\zeta(2) = \frac{1}{6\cdot 8^2} , \tag{24}$$

whilst the latter integral can be accurately evaluated using the trapezium formula on the [0, 3] interval. Having confined the range of $k$ to the $[\mu, 3]$ interval, accordingly, it is actually enough to restrict oneself to $0 \leq l, n \leq$ 10–15 when computing $L(n), L$. We took $0 \leq l \leq 14$, $0 \leq n \leq 10$. Since the Wronskian of (11) does not depend on $x$, the latter should be chosen in the region where the potentials $V^+$ are already small enough. Besides, potentials $V$ are really equal to 0 when $x <$-12.5. We computed $\Gamma(k,l,n)$ according to (11) at $x = 500$, where $f^{\pm}$ were obtained from the Volterra integral equations, respectively, (12) and (16). For this aim, according to the methods developed for numerical solutions of integral equations (see e. g., Ref. [8]), those of (12) and (16) have been replaced by systems of linear algebraic equations which can be gained when calculating the conforming integrals by the trapezium formula, respectively, for the [-12.5, 500] and [500, 600] intervals. The sought values of $f^{\pm}$ were obtained as the solutions to the above linear systems. After this the derivatives $(f^{\pm})'_x$ were evaluated in accordance with (13) and (17) while employing the values obtained for $f^{\pm}$. The typical behaviour of $\Gamma(k,l,n)$, for instance, for charged kaons is presented in Fig. 2.

Fig. 2. Typical behavior of barrier transparencies for charged kaons.

As an illustration Fig. 3 represents the untwisted $L(0)/A$ and the all configurations $L/A$ luminosities with $A$ of (9) for charged kaons as functions of $k$ while Fig. 4 presents luminosity for neutral pions (no interaction with monopoles). The areas under the curves give the corresponding values of $L(0)/A$ and $L/A$.

Fig. 3. Untwisted and all configurations luminosities for charged kaons.

Fig. 4. Luminosity for neutral pions.

In Table 1 the data on computation of the untwisted and the all configurations luminosities are represented for all meson species under consideration. Finally, we can introduce the *total* untwisted $\mathfrak{L}_0$ (as the sum over all species in the second column of Table 1) and the *total* $\mathfrak{L}$ (as the sum over all species in the third column of Table 1) luminosities multiplied by $A$ that should be detected by an external observer near black hole. We shall have

$$\mathfrak{L}_0 = 0.207299 \cdot 10^{23} \text{ erg} \cdot \text{s}^{-1}, \mathfrak{L} = 0.233520 \cdot 10^{23} \text{ erg} \cdot \text{s}^{-1}$$

so that contribution owing to Dirac monopoles amounts to 11.2284 % of $\mathfrak{L}$.

Table 1
Untwisted and the all configurations luminosities

| Particle | $L(0)/A$ | $L/A$ | Monopole contribution of $L$ (%) |
|---|---|---|---|
| $\pi^0$ | $0.938559 \cdot 10^{-4}$ | $0.938559 \cdot 10^{-4}$ | 0 |
| $\pi^+$ | $0.110489 \cdot 10^{-3}$ | $0.134436 \cdot 10^{-3}$ | 17.8124 |
| $\pi^-$ | $0.110489 \cdot 10^{-3}$ | $0.134436 \cdot 10^{-3}$ | 17.8124 |
| $K^0$ | $0.110660 \cdot 10^{-3}$ | $0.110660 \cdot 10^{-3}$ | 0 |
| $\overline{K}^0$ | $0.110660 \cdot 10^{-3}$ | $0.110660 \cdot 10^{-3}$ | 0 |
| $K^+$ | $0.110658 \cdot 10^{-3}$ | $0.134616 \cdot 10^{-3}$ | 17.7979 |
| $K^-$ | $0.110658 \cdot 10^{-3}$ | $0.134616 \cdot 10^{-3}$ | 17.7979 |

## 4 Concluding remarks

Although our algorithm works well only for small $\mu$, it seems to us that including heavier mesons or considering more massive black holes will not essentially alter the results of the paper, so long as in either case $\mu$ will increase but it is clear that luminosity $L \to 0$ at $\mu \to +\infty$ and, as a consequence, the contributions from heavier mesons will be far smaller than the ones for pions and kaons. The more interesting case is that of spinor field which describes genuine elementary particles (electrons, neutrinos etc.) and, as has recently been shown [9], where also there exist TICs on black holes due to interaction with Dirac monopoles. We hope to shed some light on the spinor case elsewhere.

## 5 Acknowledgements

The work of Goncharov was supported in part by the Russian Foundation for Basic Research (grant No. 98-02-18380-a) and by GRACENAS (grant No. 6-18-1997).

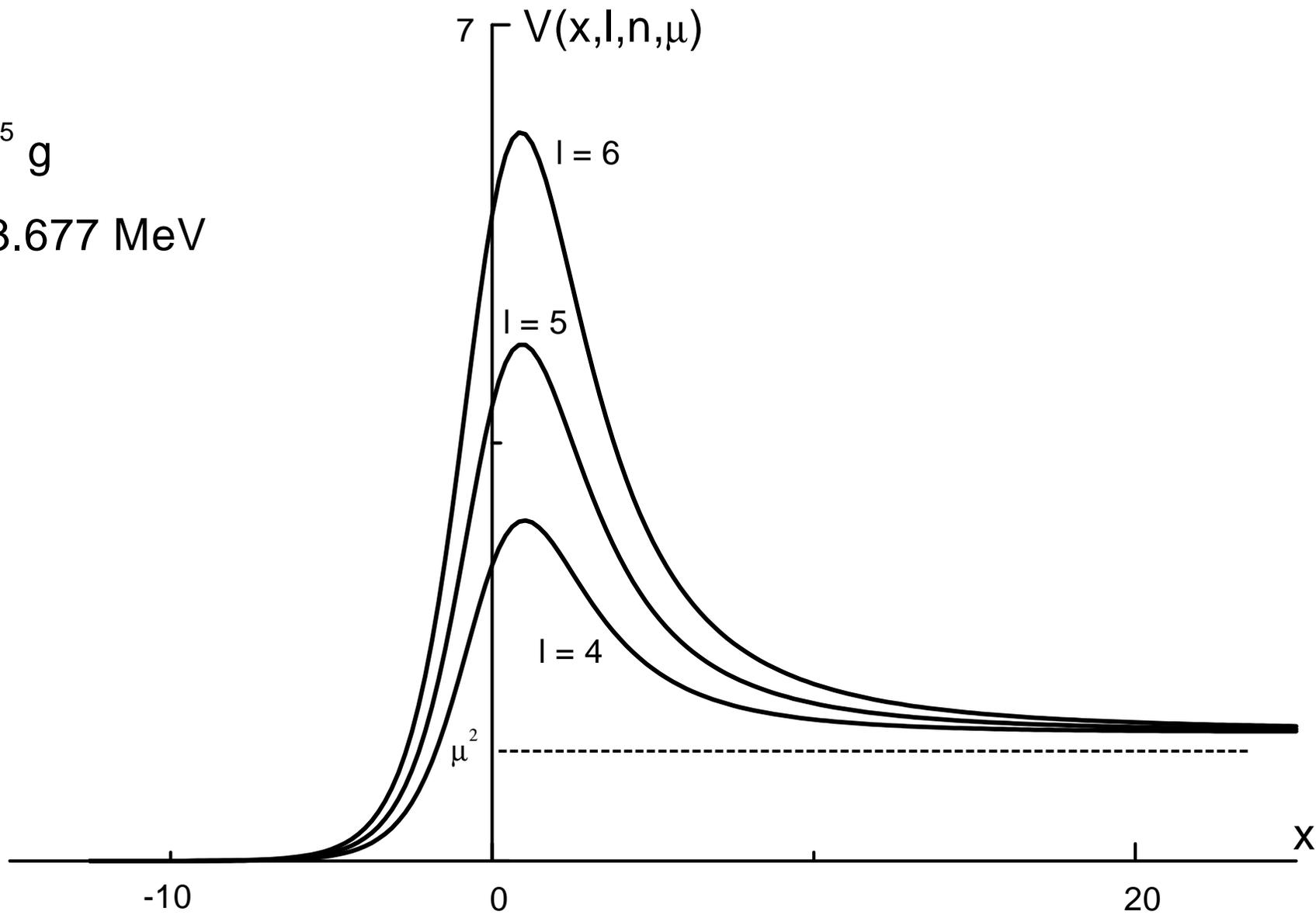

Fig. 1

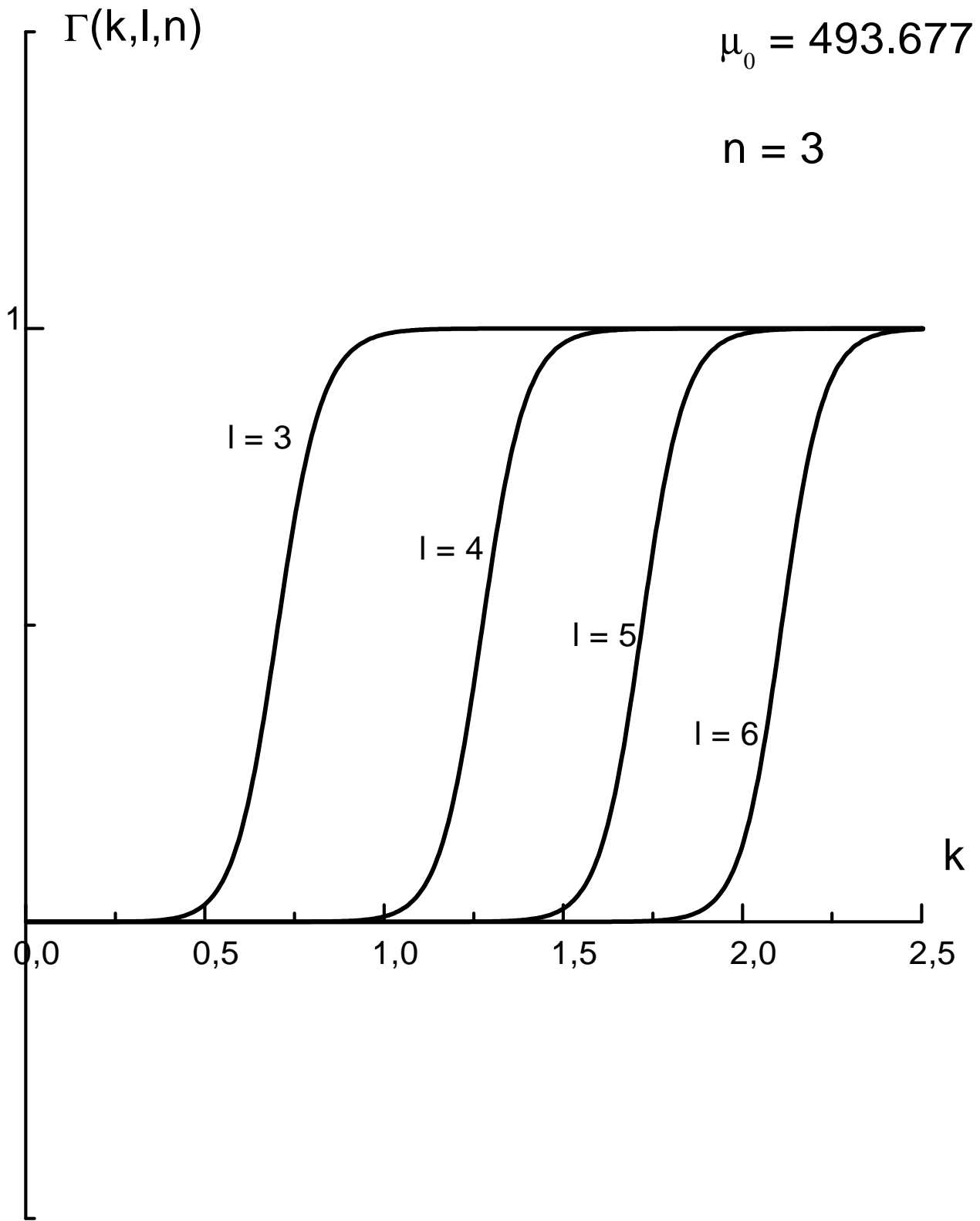

Fig. 2

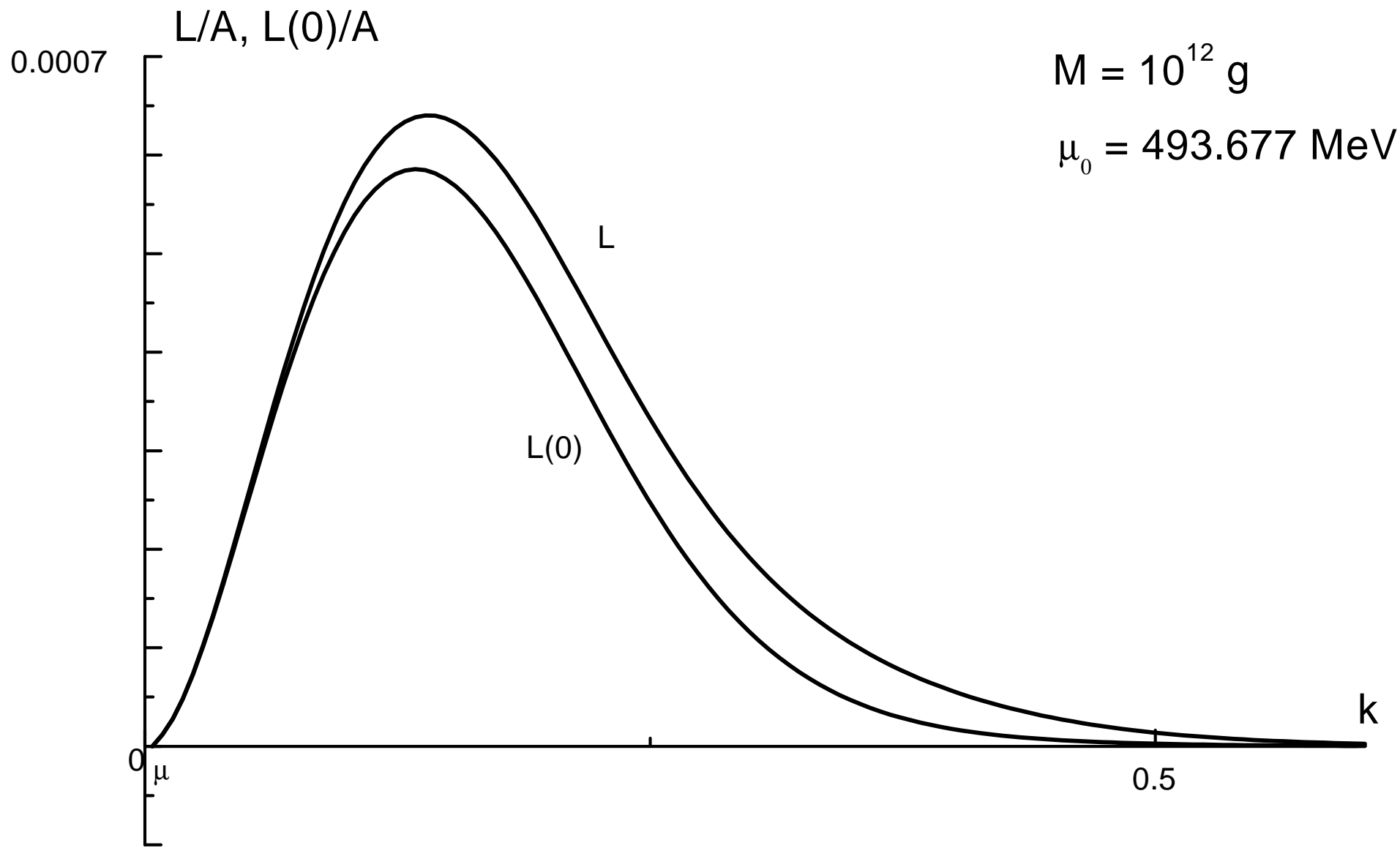

Fig. 3

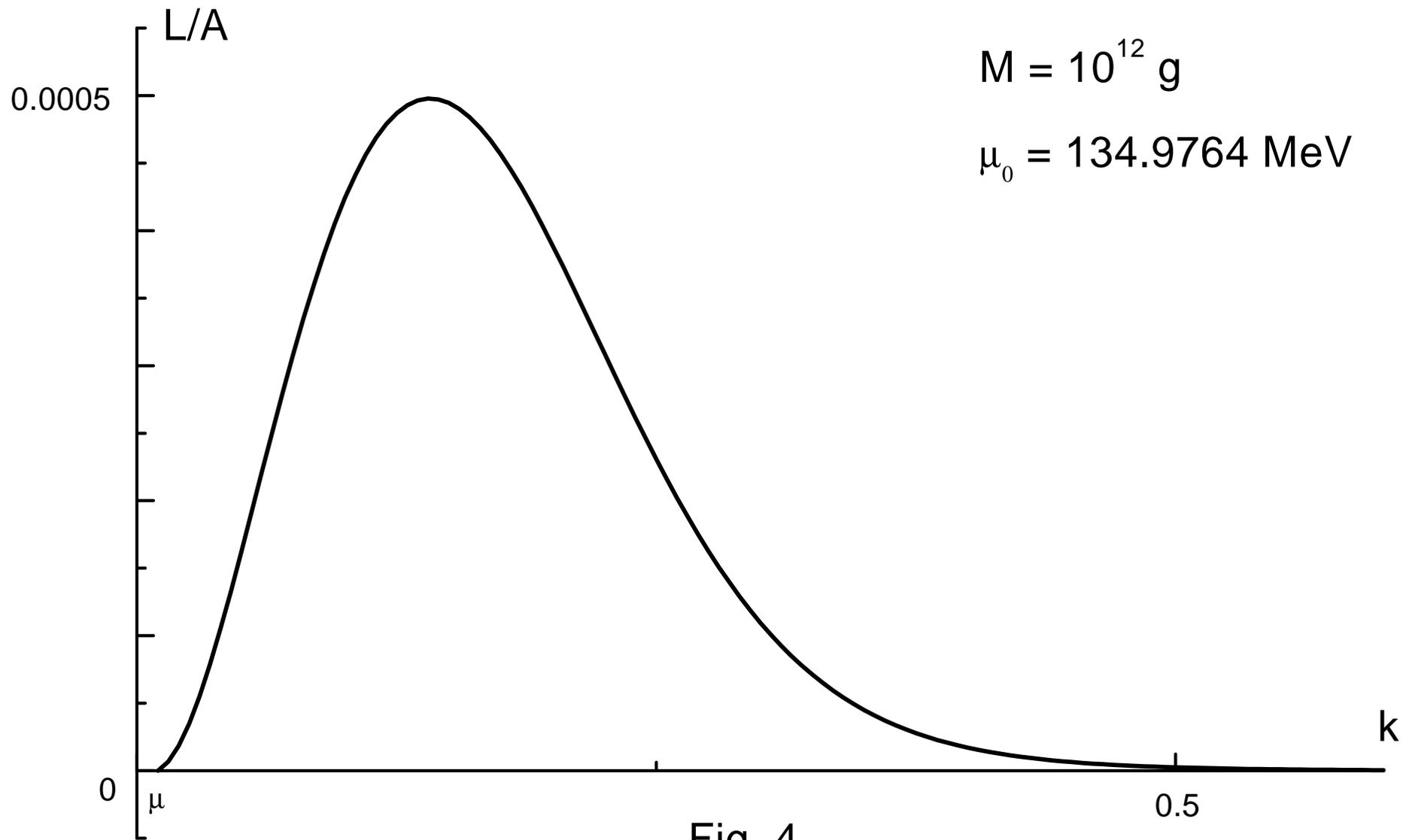

Fig. 4